\documentclass[numberedappendix,useAMS,usenatbib,letterpaper]{mn2e}
\usepackage[pass]{geometry}
\usepackage{amssymb,amsmath}
\usepackage{verbatim}
\usepackage{color}
\usepackage{hyperref}
\usepackage{booktabs}
\usepackage{needspace}
\usepackage{enumitem}
\usepackage{subfig}
\usepackage{graphicx}
\definecolor{linkcolor}{rgb}{0,0,0.25}
\hypersetup{
  colorlinks=true,        
  linkcolor=linkcolor,    
  citecolor=linkcolor,    
  filecolor=linkcolor,    
  urlcolor=linkcolor      
}
\newcounter{address}
\DeclareMathAlphabet{\mathsc}{OT1}{cmr}{m}{sc}
\def\testbx{bx}%
\DeclareRobustCommand{\ion}[2]{%
\relax\ifmmode
\ifx\testbx\f@series
{\mathbf{#1\,\mathsc{#2}}}\else
{\mathrm{#1\,\mathsc{#2}}}\fi
\else\textup{#1\,{\mdseries\textsc{#2}}}%
\fi}

\title[Galactic structure with guiding centres]
{The power of co-ordinate transformations in dynamical interpretations of Galactic structure}

\author[J. A. S. Hunt et al.]
{\parbox{\textwidth}{Jason A.~S.~Hunt$^1$, Kathryn V. Johnston$^{2,1}$, Alex R. Pettitt$^3$ \\ Emily C. Cunningham$^1$, Daisuke Kawata$^4$ and David W. Hogg$^{1,5,6}$}%
\\
\\
$^{1}$ Center for Computational Astrophysics, Flatiron Institute, 162 5th Av., New York City, NY 10010, USA\\
$^2$ Department of Astronomy, Columbia University, New York, NY 10027, USA\\
$^3$ Department of Physics, Faculty of Science, Hokkaido University, Sapporo 060-0810, Japan\\
$^4$ Mullard Space Science Laboratory, University College London, Holmbury St. Mary, Dorking, Surrey, RH5 6NT, UK\\
$^5$ Center for Cosmology and Particle Physics, Department of Physics, New York University, 726 Broadway, New York, NY 10003, USA\\
$^6$ Max-Planck-Institut f{\"u}r Astronomie, K{\"o}nigstuhl 17, D-69117 Heidelberg, Germany
}

\pagerange{\pageref{firstpage}--\pageref{lastpage}}
\pubyear{2019}

\begin{document}

\maketitle

\label{firstpage}

\begin{abstract}
$Gaia$ DR2 has provided an unprecedented wealth of information about the positions and motions of stars in our Galaxy, and has highlighted the degree of disequilibria in the disc. As we collect data over a wider area of the disc it becomes increasingly appealing to start analysing stellar actions and angles, which specifically label orbit space, instead of their current phase space location. Conceptually, while $\bar{x}$ and $\bar{v}$ tell us about the potential and local interactions, grouping in action puts together stars that have similar frequencies and hence similar responses to dynamical effects occurring over several orbits. Grouping in actions and angles refines this further to isolate stars which are travelling together through space and hence have shared histories. Mixing these coordinate systems can confuse the interpretation. For example, it has been suggested that by moving stars to their guiding radius, the Milky Way spiral structure is visible as ridge-like overdensities in the $Gaia$ data \citep{Khoperskov+19b}. However, in this work, we show that these features are in fact the known kinematic moving groups, both in the $L_z-\phi$ and the $v_{\mathrm{R}}-v_{\phi}$ planes. Using simulations we show how this distinction will become even more important as we move to a global view of the Milky Way. As an example, we show that the radial velocity wave seen in the Galactic disc in $Gaia$ and APOGEE should become stronger in the action-angle frame, and that it can be reproduced by transient spiral structure.
\end{abstract}


\section{Introduction}\label{intro}
The Milky Way is known to be a barred spiral galaxy \citep[e.g.][]{BS91,W92}, but it remains challenging to obtain a global picture of our Galaxy owing to our position within, and the complicated observational selection effects present in the data from any survey, such as imposed by the dust extinction. However, recent surveys such as the European Space Agency (ESA)'s $Gaia$ mission \citep{GaiaMission} and the Sloan Digital Sky Survey (SDSS) Apache Point Observatory Galactic Evolution Experiment \citep[APOGEE;][]{MAPOGEE17} are revealing more and more of the structure of the Galaxy we live in, and the future is bright with many upcoming surveys such as the SDSS-V Milky Way Mapper \citep[e.g.][]{Kollmeier+19} and the Legacy Survey of Space and Time \citep[LSST;][]{LSST} at the Vera Rubin Observatory.

We are now at the stage where we can start to make large scale maps of the Milky Way structure and kinematics from the stellar data alone \citep[e.g.][]{GCKatz+18,KBCCGHS18,Antoja+18,Anders+19,BLHM+19,Eilers+20}, allowing us to trace specific features for many kpc across the disc. For example, it is now clear that the Solar neighbourhood moving groups are not merely local features, but are instead the local projection of large scale kinematic structure that extend for many kpc across the disc. However, even with this new perspective, the literature is yet to converge on their origin.

In recent years, measurements of the bar's length and pattern speed appear to be converging on a long bar \citep[e.g.][]{WGP15,Clarke+19}, with a half-length of around 5 kpc and a pattern speed of around $40\pm3$ km s$^{-1}$ kpc$^{-1}$ \citep[e.g.][]{2017MNRAS.465.1621P,SSE19,BLHM+19}, replacing the older picture of a short fast bar \citep[e.g.][]{D00}. A long bar with such a pattern speed should have an impact on the Solar neighbourhood kinematics, most likely through the Corotation Resonance \citep[CR; e.g][]{P-VPWG17} and the Outer Lindblad Resonance (OLR), and potentially higher order resonances depending on the structure of the bar \citep[e.g.][]{HB18,MFSWG19,Asano+20}. However, even a few km s$^{-1}$, e.g. the current level of uncertainty, is enough to make it unclear which feature in the Solar neighbourhood kinematics arises from which resonance, and if the bar is slowing then multiple features can be explained by the resonant sweeping of a single resonance \citep{CFS19}.


The picture is further complicated by the Milky Way spiral structure. For example, in \cite{Hunt+19} we showed that transient winding spiral arms can reproduce the Solar neighbourhood kinematics in combination with a variety of bar models. It is non-trivial to disentangle the signatures of the Galactic bar and spiral structure, especially when the number of spiral arms and nature of the spiral structure itself remain uncertain. Similarly, \citet{Pettitt2020} showed that global features in velocity space unveiled by \emph{Gaia} could be equally well reproduced by spiral or bar features.

To date, most methods of locating the Milky Way spiral structure rely on observations of gas, and young star forming regions that are thought to be associated with the spiral arms \citep[e.g.][]{DHT01,LBH06,Retal14,HH14}. However, different theories of spiral arm formation and evolution predict differences in whether there is an offset in the location of the spiral density enhancement between the stellar and gaseous component \citep[e.g.][]{Baba2015}. 

Thus, it is important to also locate the spiral structure in the stellar component of the Milky Way, both to reinforce or contradict the maps based on gas or Masers, but also to test the underlying theory of spiral arm dynamics. A few studies have found evidence for spiral arms in the stellar number counts \citep[e.g.][]{Poggio+2018,Chen+19,Miyachi+19} but this is not an easy task owing to the high level of dust extinction in the disc plane, and the difficulty of obtaining a complete sample across many kpc.

With the increasing quantity and quality of data, it becomes increasingly appealing to start analysing the orbit structure of the Milky Way, using actions and angles (i.e., orbit labels) instead of positions and velocities alone. $Gaia$ DR2 \citep{DR2} has now been well explored in action space, both in terms of the data \citep{TCR18}, and with comparison to various models for Galactic bar and spiral structure \citep[e.g.][]{STCCR19,Hunt+19,Trick+19} in attempts to link the numerous kinematic features present in the observations with the mechanism which causes them.


\begin{figure}
    \centering
    \includegraphics[width=\hsize]{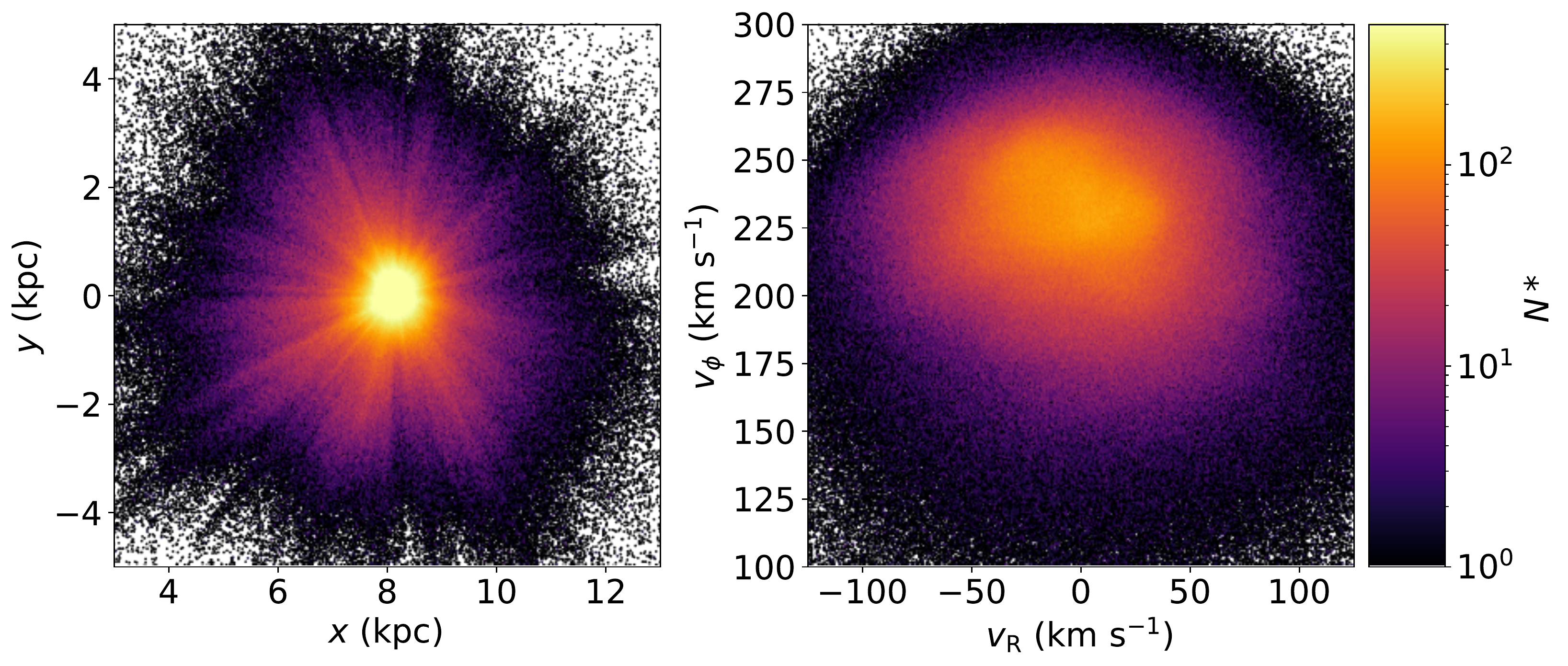}
    \caption{Logarithmic number density for the $x-y$ plane (left) and the $v_{\mathrm{R}}-v_{\phi}$ plane (right) for the $Gaia$ sample as described in Section \ref{data}, shown only for comparison.}
    \label{XRGYRG}
\end{figure}

However, while actions and angles remain an attractive method of examining galactic structure and kinematics, care must be taken when working in these coordinate systems, or in particular when mixing them with other canonical coordinate systems. It is important to be aware of the physical meaning of such transformations, and also the significant impact that observational selection functions can have when working with real data. This is not a new revelation, and the limitations imposed by the selection function when interpreting local dynamical signatures, especially when employing actions and angles, has been shown in numerous earlier works \citep[e.g.][]{Sellwood2010,McMillan2011}. While $Gaia$ DR2 allows us to see beyond the Solar neighbourhood the sample is still dominated by local objects, and local selection effects remain important \citep[e.g.][]{STCCR19}.

In this work, we intend to illustrate the utility of the transformation from physical space to action-angle space, discuss for what questions it is desirable to use such a framework, and comment on the hazards of mixing canonical conjugate coordinate systems. For specific examples, we show how the kinematic response to the Galactic potential can be revealed and dissected with a transformation to action angle space, both in the context of a local sample from $Gaia$ DR2, and in the study of Milky Way spiral structure further across the disc. In Section \ref{data} we define our coordinate systems and use $Gaia$ DR2 data to illustrate their properties. In Section \ref{simulation} we use dynamical simulations to illustrate what the different coordinate spaces will tell us when we have more complete maps of the disc if we can also fully account for the selection function. In Section \ref{wave} we use our coordinates to show a link between the spiral structure and the radial velocity wave observed in the Solar neighbourhood kinematics. In Section \ref{summary} we give our conclusions.

\section{The Solar neighbourhood as seen by $Gaia$ DR2}\label{data}
In this Section, we show data from $Gaia$ DR2 in different coordinate systems. For our primary sample, we first select all stars with a measured radial velocity and then apply the quality cuts suggested in \cite{SME19}, namely stars with a color of $G_{\mathrm{BP}}-G_{\mathrm{RP}}<1.5$, a magnitude of $G<14.5$, a fractional parallax error of $\pi/\sigma_{\pi}>4$, a parallax uncertainty cut of $\sigma_{\pi}<0.1$, a BP-RP excess flux factor of $1.172<E_{\mathrm{BPRP}}<1.3$, and with more than 5 visibility periods used. For the results in this paper we calculate distances by naively inverting the parallax, $d=1/\pi$. However, we repeated the analysis with the Bayesian distance estimates of \cite{SME19} (using the \sc{gaiaRVdelp54delsp43 }\rm sample), and confirmed the resulting kinematic structure to be the same. 

\subsection{$Gaia$ DR2 in projected physical space $(x,y)$}\label{cart}
Firstly, as an illustration and purely for comparison with the below sections we show the distribution of stars in the Solar neighbourhood and the planar kinematics. This is not new and can be seen in many $Gaia$ DR2 publications. We assume a distance to the Galactic centre of $R_0=8.178$ kpc \citep{Gravity+19}, and the Sun's height about the disc plane as 20.8 pc \citep{BB19}. We calculate the vertical and azimuthal Solar motion by combining $R_0$ with the proper motion measurement of Sgr A* of $(\mu_l,\mu_b)=(-6.411\pm0.008,-0.219\pm0.007)$ \citep{RB20}. Thus we have $v_{\odot}=248.5$ km s$^{-1}$ and $w_{\odot}=8.5$ km s$^{-1}$.

Fig. \ref{XRGYRG} shows the logarithmic number density for the $x-y$ plane (left) and the $v_{\mathrm{R}}-v_{\phi}$ plane (right) for the sample described in the previous section. The left panel shows that most of the sample are within around 2 kpc from the Sun, with obvious extinction features resulting in significant incompleteness. The right hand panel shows very little structure in the $v_{\mathrm{R}}-v_{\phi}$ plane. This is unsurprising, both from the rather lax quality cuts, but mainly because kinematic structure is not expected to be consistent over such a large spatial range, i.e. $v_{\mathrm{R}}$ and $v_{\phi}$ are only equivalent to orbit labels at a single point.

\begin{figure}
    \centering
    \includegraphics[clip, trim=8.7cm 4cm 11.5cm 2cm, width=\hsize]{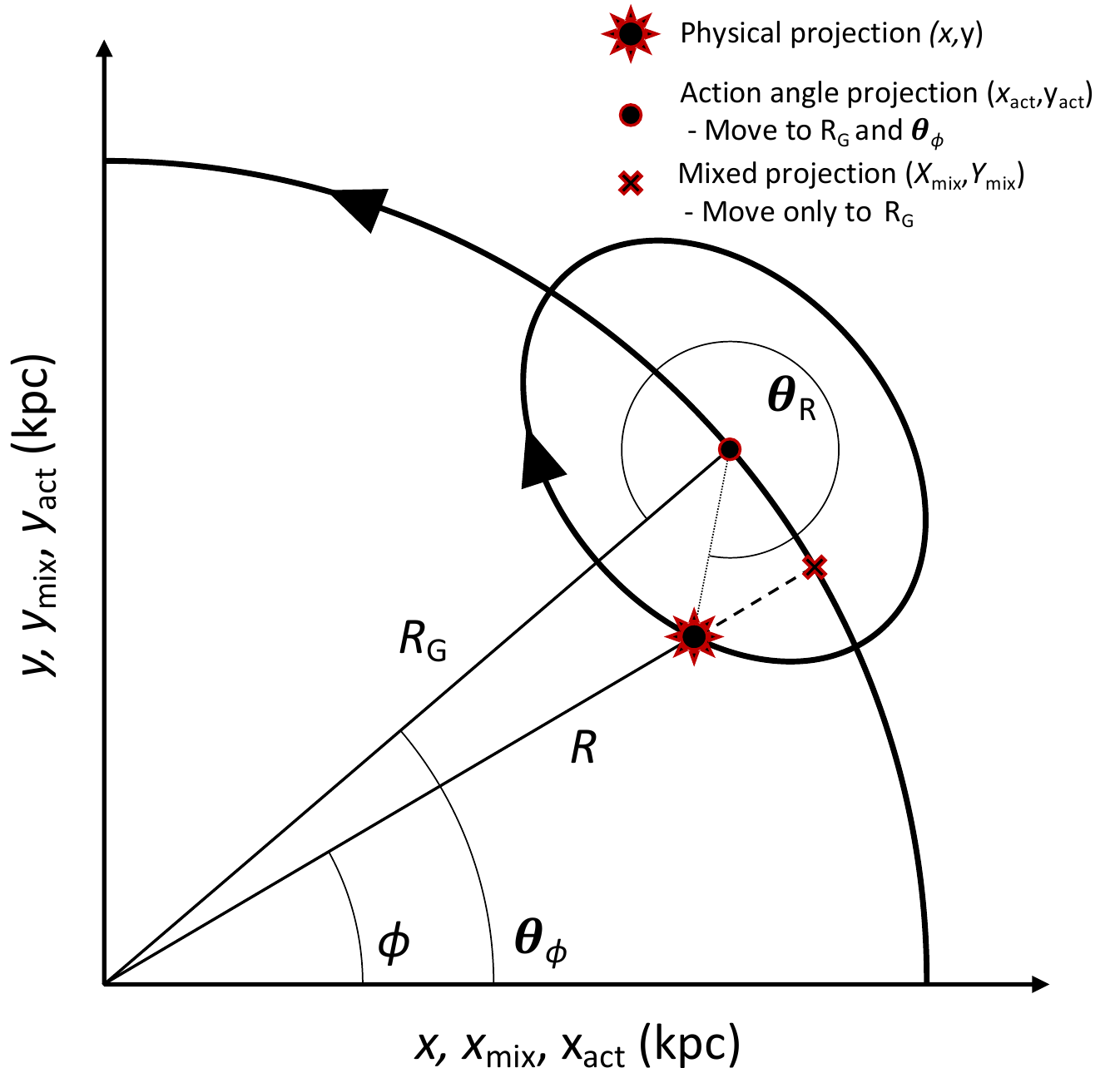}
    \caption{Diagram of planar action angle coordinates showing the relation between a stars current distance to the galactic centre, $R$, and azimuth $\phi$, with the guiding centre radius, $R_{\mathrm{G}}$, azimuthal angle $\theta_{\phi}$ and radial angle, $\theta_{\mathrm{R}}$. The thick black lines show the motion of star around the epicycle, and along the circular orbit. Note that this diagram assumes a left-handed coordinate system as used e.g. in \sc{galpy}\rm, and in this work the Sun would lie on the $x$ axis such that $\phi=0$ and $\theta_{\phi}=0$ along the Sun-Galactic centre line.}
    \label{AAdeg}
\end{figure}

\begin{figure}
    \centering
    \includegraphics[width=\hsize]{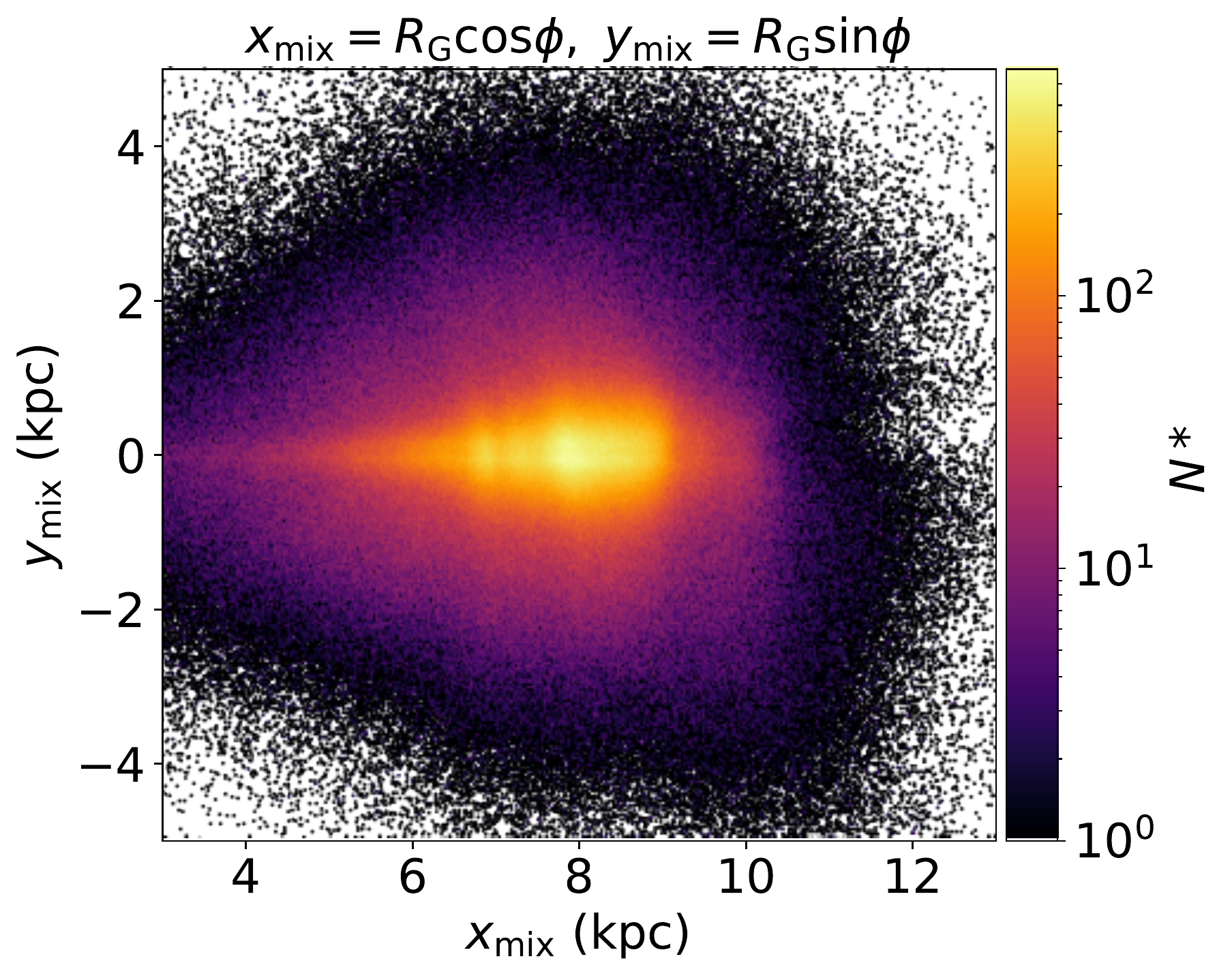}
    \includegraphics[width=\hsize]{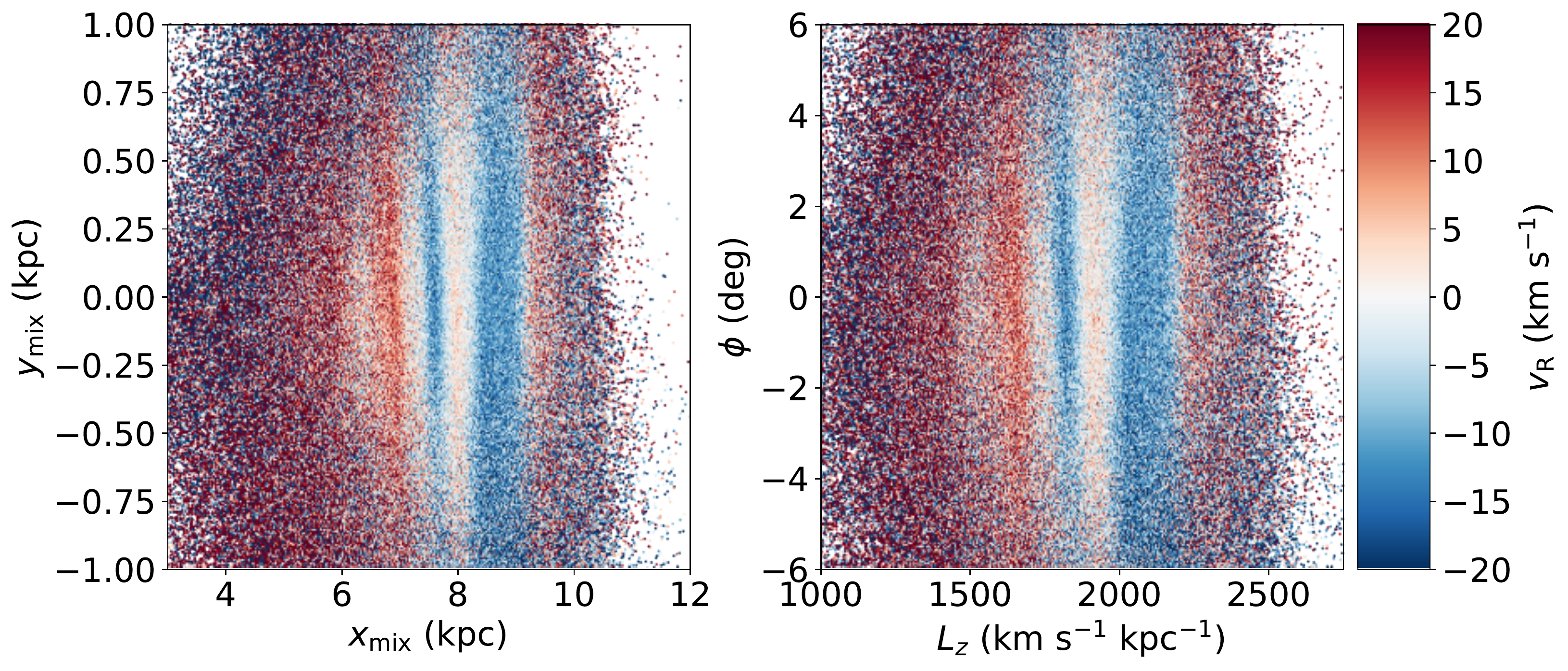}
    \caption{\textbf{Upper:} Logarithmic number density for the $x_{\mathrm{mix}}-y_{\mathrm{mix}}$ plane where stars have been moved radially to their guiding radius, but with their azimuth unchanged. \textbf{Lower:} The $x_{\mathrm{mix}}-y_{\mathrm{mix}}$ plane (left) and the $L_z-\phi$ plane (right) colored by mean radial velocity, $v_{\mathrm{R}}$ (km s$^{-1}$), over a smaller range than Fig. \ref{XRGYRG}.}
    \label{XRGYRGRv}
\end{figure}

\subsection{$Gaia$ DR2 in projected mixed space ($x_{\mathrm{mix}},y_{\mathrm{mix}}$)}\label{grad}
For structures like the Galactic disc, where stars move on near circular orbits in a potential dominated by an axisymmetric mass distribution, an epicyclic approximation to the orbits of stars provides useful intuition. The motion in the plane for a star with angular momentum, $L_z$, can be broken down into the mean motion of the guiding centre of the star superimposed with its epicyclic oscillation around this mean. The guiding centre is located at $(R,\phi)=(R_{\mathrm{G}},\theta_{\phi})$, where $R_{\mathrm{G}}$ is the guiding radius, defined as the radius of the circular orbit with the same angular momentum as the star, and $\theta_{\phi}$ is the guiding centre azimuth, the angle of the stars guiding centre with respect to the Sun-Galactic centre line. In this case, $L_z=J_{\phi}$, the azimuthal action, and the planar position is $\theta_{\phi}$. For the mixed coordinate system we approximate $R_{\mathrm{G}}=L_z/V_{\mathrm{LSR}}$, where $L_z=R\times v_{\phi}$. This effectively assumes a completely flat rotation curve. Hence the guiding radius $R_{\mathrm{G}}$ is a projection of $J_{\phi}$.

For illustration and to aid the description, Fig. \ref{AAdeg} shows a diagram of the epicycle approximation and the relation between an example star's physical location (red star) as determined by its current distance to the galactic centre, $R$, and azimuth $\phi$, with its guiding centre radius, $R_{\mathrm{G}}$, azimuthal angle $\theta_{\phi}$ and radial angle, $\theta_{\mathrm{R}}$. The thick black lines show the motion of star around the epicycle, and along the circular orbit. The ratio of the major and minor axis of the epicycle illustrated here is $\sqrt{2}$, which is appropriate for a flat rotation curve. Note that this diagram assumes a left-handed coordinate system as used e.g. in \sc{galpy }\rm \citep{B15}, whereas the commonly used \sc{astropy }\rm library assumes a right-handed coordinate system. Note also that the epicyclic motion is retrograde compared to the motion of the guiding centre, i.e. the epicyclic motion is clockwise for an anti-clockwise orbit. 

Following \cite{Khoperskov+19b} we define mixed Cartesian coordinates $(x_{\mathrm{mix}},y_{\mathrm{mix}})$ as the frame where stars positions are shifted radially to their guiding radius, $R_{\mathrm{G}}$, without altering their azimuth. We set $x_{\mathrm{mix}}=R_{\mathrm{G}}\cos(\phi)$ and $y_{\mathrm{mix}}=R_{\mathrm{G}}\sin(\phi)$. In Fig. \ref{AAdeg} this transformation is represented by the red cross for the example star.

This is essentially a half transformation to action-angle space, i.e. we are now using the azimuthal action, $J_{\phi}=L_z$ in combination with the Cartesian angle $\phi$. This mix of coordinate systems will give us a new perspective on the data, but note that the physical meaning of such a space is a mix of current physical position, with an orbit label. Thus, we call these mixed coordinates.

\begin{figure*}
    \centering
    \includegraphics[clip, trim=3cm 0cm 0cm 0cm, width=\hsize]{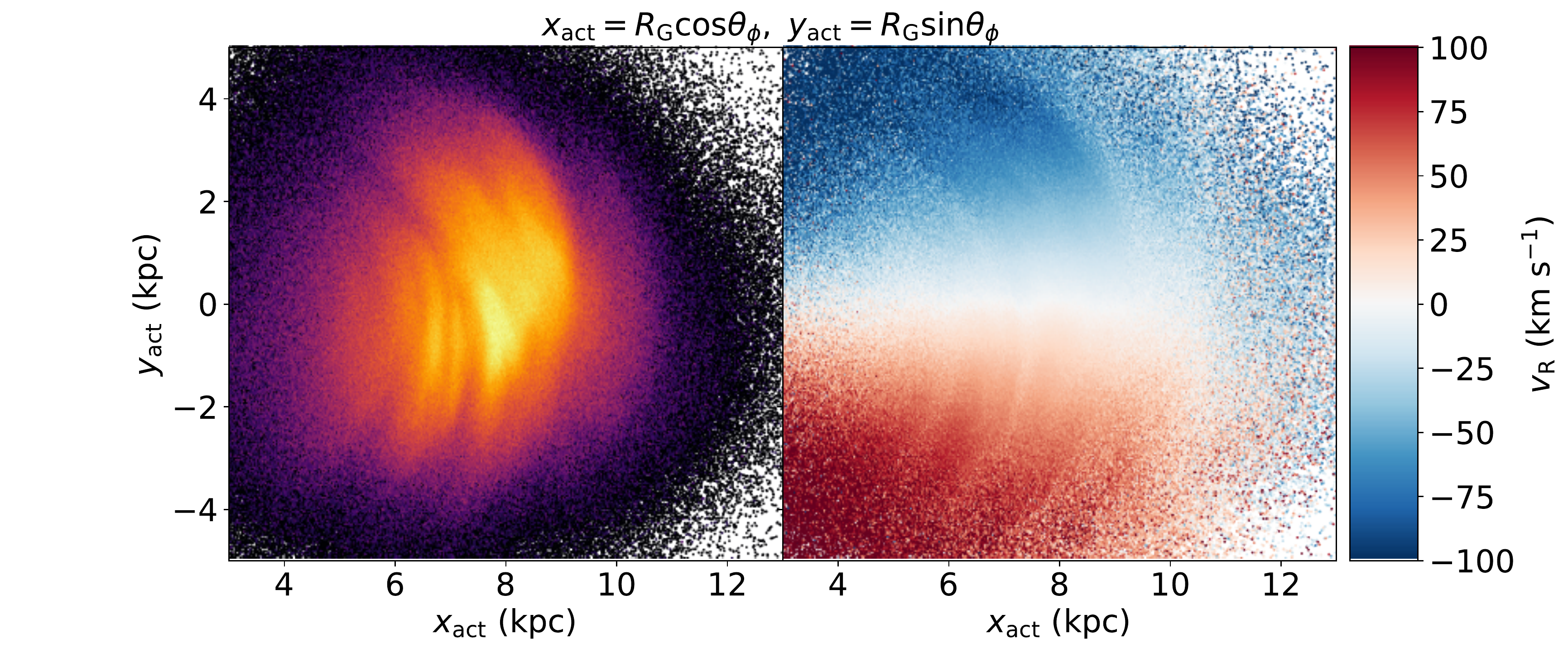}
    \caption{Logarithmic number density (left) and radial velocity map for the $Gaia$ DR2 sample in guiding centre Cartesian coordinates, $x_{\mathrm{act}}-y_{\mathrm{act}}$, i.e. where stars have been moved both radially and azimuthally to their guiding centre.}
    \label{XGYG}
\end{figure*}

The top panel of Fig. \ref{XRGYRGRv} shows the logarithmic number density of the $Gaia$ sample in the $x_{\mathrm{mix}}-y_{\mathrm{mix}}$ plane. The same ridge-like features are visible as in Fig. 1 of \cite{Khoperskov+19b}, although without their sampling and unsharp mask the pattern is significantly reduced away from the $y_{\mathrm{mix}}=0$ line. They interpret these ridges as evidence of spiral structure in the Milky Way.

The lower left panel of Fig. \ref{XRGYRGRv} shows the $x_{\mathrm{mix}}-y_{\mathrm{mix}}$ plane over a smaller $y_{\mathrm{mix}}$ range, colored by radial velocity, $v_{\mathrm{R}}$ (km s$^{-1}$). For a small value of the angle, $\phi$, or a small range of $y_{\mathrm{mix}}$, this is essentially the $L_z-\phi$ plane, which is shown in the right panel of Fig. \ref{XRGYRGRv}. 
For example, via the small angle approximation, assuming the Sun is located at $(x,y)=(8.178,0)$ kpc, with $\phi=0$ deg, $x_{\mathrm{mix}}=R_{\mathrm{G}}\cos(\phi)\approx R_{\mathrm{G}}$, and with $R_{\mathrm{G}}=L_z/V_{\mathrm{circ}}$ for a constant $V_{\mathrm{circ}}$, $R_{\mathrm{G}}$ is a scaled angular momentum, thus $x_{\mathrm{mix}}\propto L_z$. Similarly, by the small angle approximation, $y_{\mathrm{mix}}=R_{\mathrm{G}}\sin(\phi)\propto\phi$ (with a small $L_z$ dependence). 


The $L_z-\phi$ plane is already well explored and the reader should see \cite{FS19} for a more detailed analysis. The lower panels of Fig. \ref{XRGYRGRv} show that these different ridges have distinct radial velocity signatures, which are in agreement with those already shown in \cite{FS19}. The sampling scheme and unsharp mask employed in \cite{Khoperskov+19b} help extend the visibility of these features further from the Sun, but we suggest that they are previously known kinematic signatures, not coherent structure in physical space such as spiral arms. 


\begin{figure}%
    \centering
    \includegraphics[clip, trim=10.5cm 3cm 13cm 1.5cm, width=\hsize]{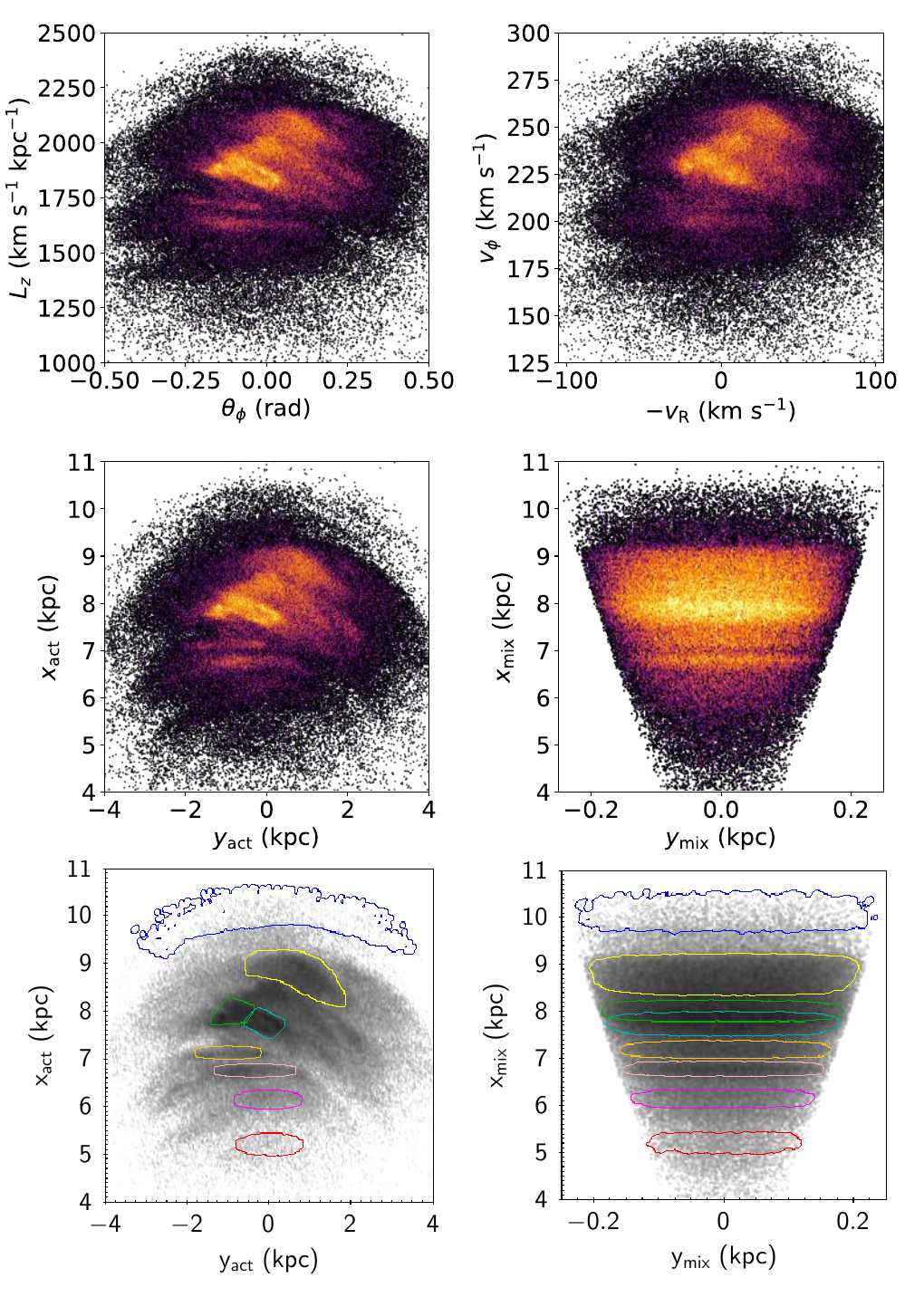}
    \caption{Logarithmic number density for the $Gaia$ DR2 sample within 150 pc in $L_z-\theta_{\phi}$ (top left), $v_R-v_{\phi}$ (top right), $y_{\mathrm{act}}-x_{\mathrm{act}}$ (middle left), and $y_{\mathrm{mix}}-x_{\mathrm{mix}}$ (middle right). An approximate selection of these moving groups in the $y_{\mathrm{act}}-x_{\mathrm{act}}$ plane (lower left) and the $y_{\mathrm{mix}}-x_{\mathrm{mix}}$ plane (lower right) for the same sample within 150 pc. Arcturus is in red, the components of Hercules are shown in bright pink, peach and orange, Pleiades is in cyan, Hyades is in green, Sirius is in yellow and the hat is in blue. This is not a rigorous selection of group members, merely an approximate selection in the left panel, which is projected into the right panel as an illustration of the relation between coordinate spaces.}%
    \label{200pc}%
\end{figure}

However, while these ridges are not the physical location of the spiral structure, there is no reason you cannot examine such a space. Actions are orbit labels, which will group stars with shared orbital history, and such an orbit map may hold information on the history of the disc, or the nature of spiral structure. Yet in this coordinate system we are only using half of the orbit information, angles are also part of the orbit label, and using both will provide more information than actions alone.

\subsection{$Gaia$ DR2 in projected action angle space ($x_{\mathrm{act}},y_{\mathrm{act}}$)}\label{gcen}
To complete the transformation to action-angle coordinates, we define guiding centre Cartesian coordinates as the frame where we shift the positions of the stars to their guiding centre azimuth, $\theta_{\phi}$, as well as the shift to $R_{\mathrm{G}}$. With this transformation, $x_{\mathrm{act}}=R_{\mathrm{G}}\cos(\theta_{\phi})$ and $y_{\mathrm{act}}=R_{\mathrm{G}}\sin(\theta_{\phi})$. In Fig. \ref{AAdeg} this transformation is represented by the red circle for the example star. In this instance we calculate $L_z$ and $\theta_{\phi}$ using the \texttt{actionAngleStaeckel} \citep{B12-2} function in \texttt{galpy}, assuming the \texttt{MWPotential2014} potential, which is fit to various observational constraints \citep{B15}. We calculate the delta parameter, which is the focal-length parameter of the prolate spheroidal coordinate system used in the approximation, using the \texttt{estimateDeltaStaeckel} routine \citep[e.g. as described in][]{2012MNRAS.426..128S}. Note that these are the axisymmetric actions and angles which are only fully conserved integrals of motion in an axisymmetric system. However, the axisymmetric approximations can still be useful when examining orbit structure in systems where the true, fully conserved integrals of motion are challenging to calculate.

Fig. \ref{XGYG} shows the $x_{\mathrm{act}}-y_{\mathrm{act}}$ plane for the $Gaia$ sample in logarithmic number density (left), and coloured by radial velocity (right). At this point the relation to the local $v_{\mathrm{R}}-v_{\phi}$ plane becomes clear \citep[e.g.][]{Dehnen98,Antoja+18}. Similar to above, for a small angle of $\theta_{\phi}$, $x_{\mathrm{act}}\propto L_z$, and in this case, for small angle, $y_{\mathrm{act}}\propto\theta_{\phi}$ (with a small $L_z$ dependence). Note that even with the relatively lenient quality cuts, the structure in the projected action-angle space is clear, as opposed to the same sample in the right hand panel of Fig. \ref{XRGYRG}. However, the observed structure is heavily influenced by the selection function. 

Over a small area of the Galaxy, such as in our sample, $\theta_{\phi}$ is a proxy for $v_{\mathrm{R}}$. I.e. stars with positive (negative) radial velocities are on the outwards (inwards) part of their epicycle and thus have their guiding centre azimuths ahead (behind) the Sun on its orbit. Thus, the right hand panel of Fig. \ref{XGYG} shows a clear relation between $\theta_{\phi}$ and $v_{\mathrm{R}}$. While this selection effect disappears if you have a complete sample across the Galactic disc (as illustrated in Section \ref{simulation}) it will continue to be a problem for future large scale survey data. Any sample of stars which extends across the disc but is dominated by local objects must be treated carefully.


\begin{figure*}
    \centering
    \includegraphics[width=\hsize]{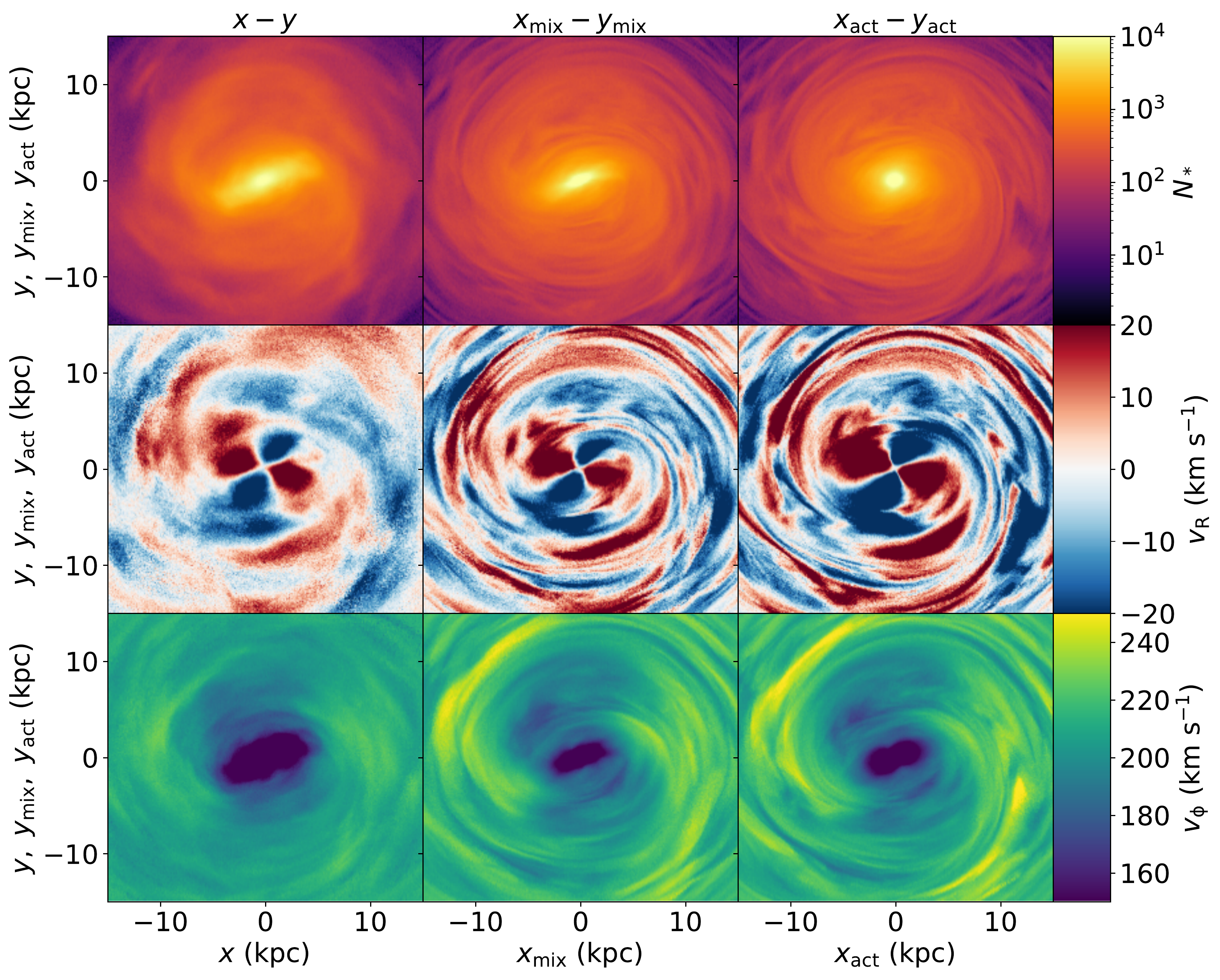}
    \caption{Face on view of model A in logarithmic number density (upper row), radial velocity (middle row) and rotation velocity (lower row) for $x-y$ (left column), $x_{\mathrm{mix}}-y_{\mathrm{mix}}$ (middle column) and $x_{\mathrm{act}}-y_{\mathrm{act}}$ (right column).}
    \label{K12}
\end{figure*}

\begin{figure*}%
    \centering
    \subfloat{{\includegraphics[width=9.5cm]{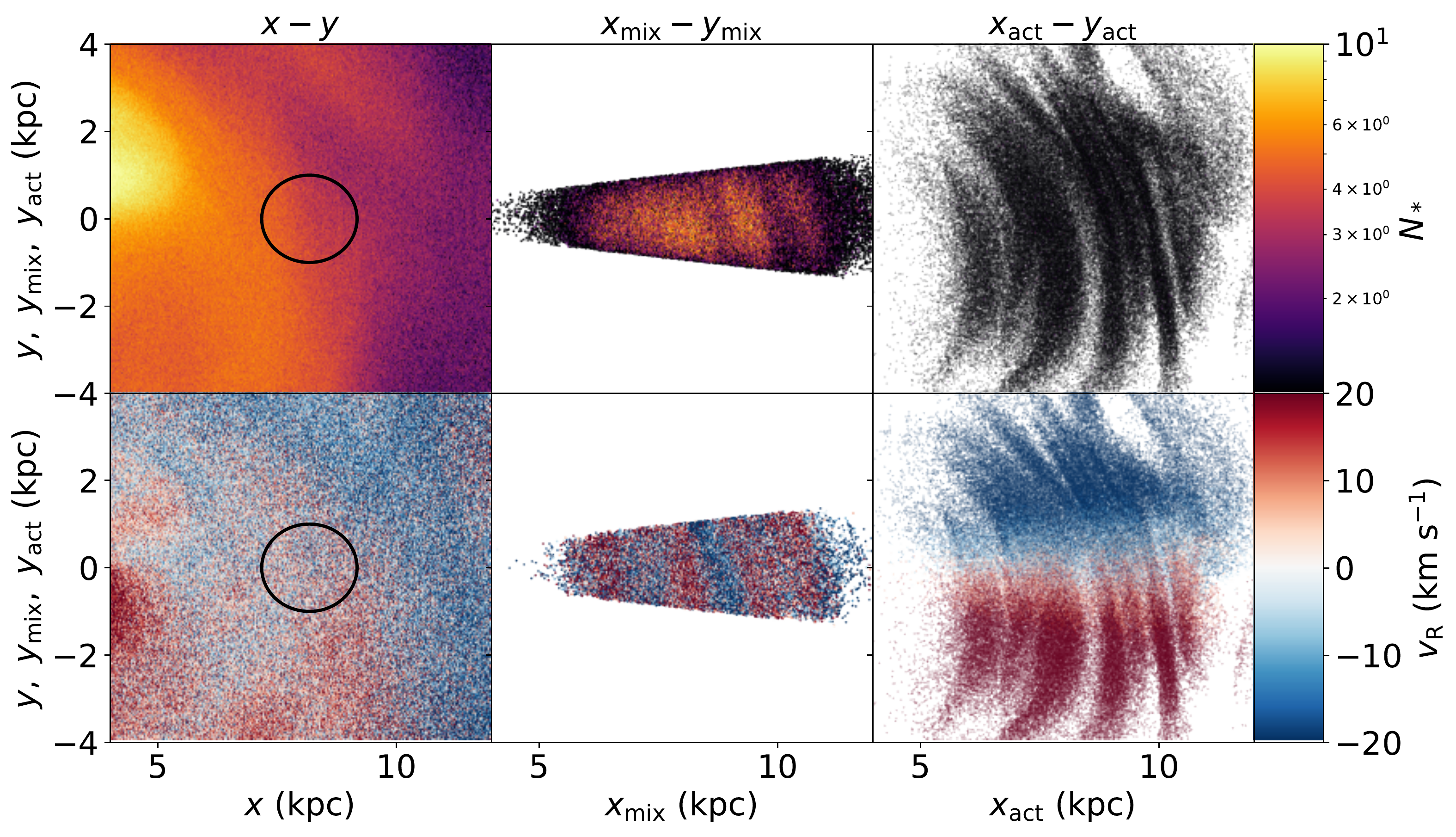}}}%
    \qquad
    \subfloat{{\includegraphics[width=7cm]{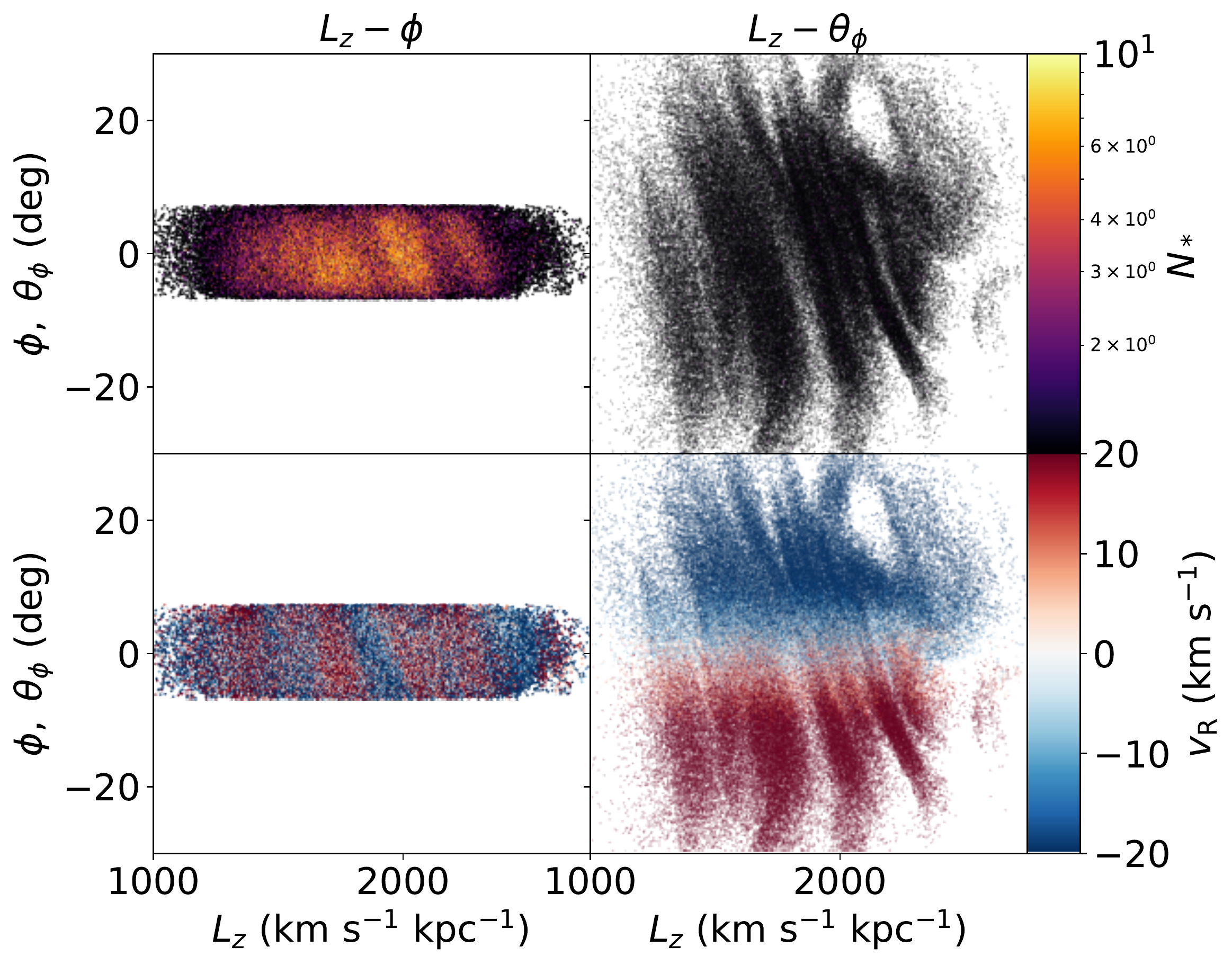}}}%
    \caption{\textbf{Left:} Sample of stars from Model A selected in a sphere of 1 kpc around $(x,y)=(8.178,0)$ (black circle) in Cartesian coordinates (left column), guiding radius coordinates (middle column) and guiding centre coordinates (right column) in number density (upper row) and colored by $v_{\mathrm{R}}$ (km s$^{-1}$; lower row). \textbf{Right:} The same sample of stars as a function of angular momentum against galactic azimuth $\phi$ (left column) and guiding centre azimuth (right column), for the number density (upper row) and colored by radial velocity (lower row). Note the similarity of the right figure with the guiding radius and guiding centre panels in the left figure.}%
    \label{sample}%
\end{figure*}

It is entirely possible (and likely) that some of this kinematic substructure arises from spiral structure \citep[e.g.][]{HHBKG18,STCCR19,Pettitt2020} but it should not be assumed that they mark the current location of spiral arms. Following this transformation we now have a map of the kinematic response to the potential, not the potential itself. As an example, the same structure in $L_z$ (for either $x_{\mathrm{mix}}-y_{\mathrm{mix}}$ or $x_{\mathrm{act}}-y_{\mathrm{act}}$) remains clear for a sample of local stars. Fig. \ref{200pc} shows a comparison of the $L_z-\theta_{\phi}$ (top left), $v_{\mathrm{R}}-v_{\phi}$ (top right) $y_{\mathrm{act}}-x_{\mathrm{act}}$ (middle left) and $y_{\mathrm{mix}}-x_{\mathrm{mix}}$ (middle right) planes for a sample of stars within 150 pc. Fig. \ref{200pc} shows that for the same sample of stars, the structure is sharper in action-angle coordinates than in the kinematics alone, even over a local area. Fig. \ref{200pc} also shows that for a local sample, $x_{\mathrm{act}}-y_{\mathrm{act}}$ is almost equivalent to $L_z-\theta_{\phi}$, but with increased curvature of the features when moving away from $y_{\mathrm{act}}=0$, purely because of the projection axes. 

The middle right panel of Fig. \ref{200pc}, shows that the structure in the $y_{\mathrm{mix}}-x_{\mathrm{mix}}$ plane is a projection of the kinematic moving groups, as seen in the other panels, smeared out in azimuth. For the list of ridges in \cite{Khoperskov+19b}, we identify SDS1 as Arcturus, SDS2 \& SDS3 as two of the three subcomponents of the Hercules stream (with the third being unresolved), SDS4 as the Hyades \& Pleiades streams, SDS5 as the Sirius stream and SDS6 as the `hat'. As a final illustration, Fig. \ref{200pc} shows an approximate selection of these moving groups in the $y_{\mathrm{act}}-x_{\mathrm{act}}$ plane (lower left) and projected into the $y_{\mathrm{mix}}-x_{\mathrm{mix}}$ plane (lower right) for the same sample within 150 pc. From bottom to top, Arcturus is in red, the components of Hercules are shown in bright pink, peach and orange, Pleiades is in cyan, Hyades is in green, Sirius is in yellow and the hat is in blue.

\section{The simulations in guiding centre space}\label{simulation}
In the previous section we reviewed the projection of $Gaia$ data into 3 coordinate systems. Our interpretation of the physical information in these coordinate systems is as follows: 1) The projected physical space, $(x,y)$, shows us the stellar density and hence the stellar component of the potential and the immediate kinematic response. 2) The projected mixed space, $(x_{\mathrm{mix}},y_{\mathrm{mix}})$, shows a mix of physical location and an orbit label. 3) The projected action angle space, ($x_{\mathrm{act}},y_{\mathrm{act}})$, shows us stars that move together in space and time, which have common reactions because of shared frequencies and a common history because of their shared phases. In this space, the response takes the longest to phase mix away. 

In this Section we examine this physical intuition using simulated galaxies with bars and spiral arms, both over a local region to compare directly to the data, and globally across the model. We do this both to illustrate what we are looking at in the $Gaia$ DR2 data with such transformations, and also to build dynamical intuition in preparation for future $Gaia$ data releases when we will be able to observe a larger fraction of the disc. We perform the same transforms to guiding radius and guiding centre coordinates as described in Sections \ref{grad} and \ref{gcen}, except using the mean circular velocity at $R=8.178$ kpc for each galaxy model. We again calculate $\theta_{\phi}$ using the \texttt{actionAngleStaeckel} \citep{B12-2} function in \texttt{galpy}, assuming \texttt{MWPotential2014}. While this is not the true galactic potential, the rotation curve is approximately correct, and the axisymmetric actions and angles are already only approximations in a non-axisymmetric system.

\begin{figure*}
    \centering
    \includegraphics[width=\hsize]{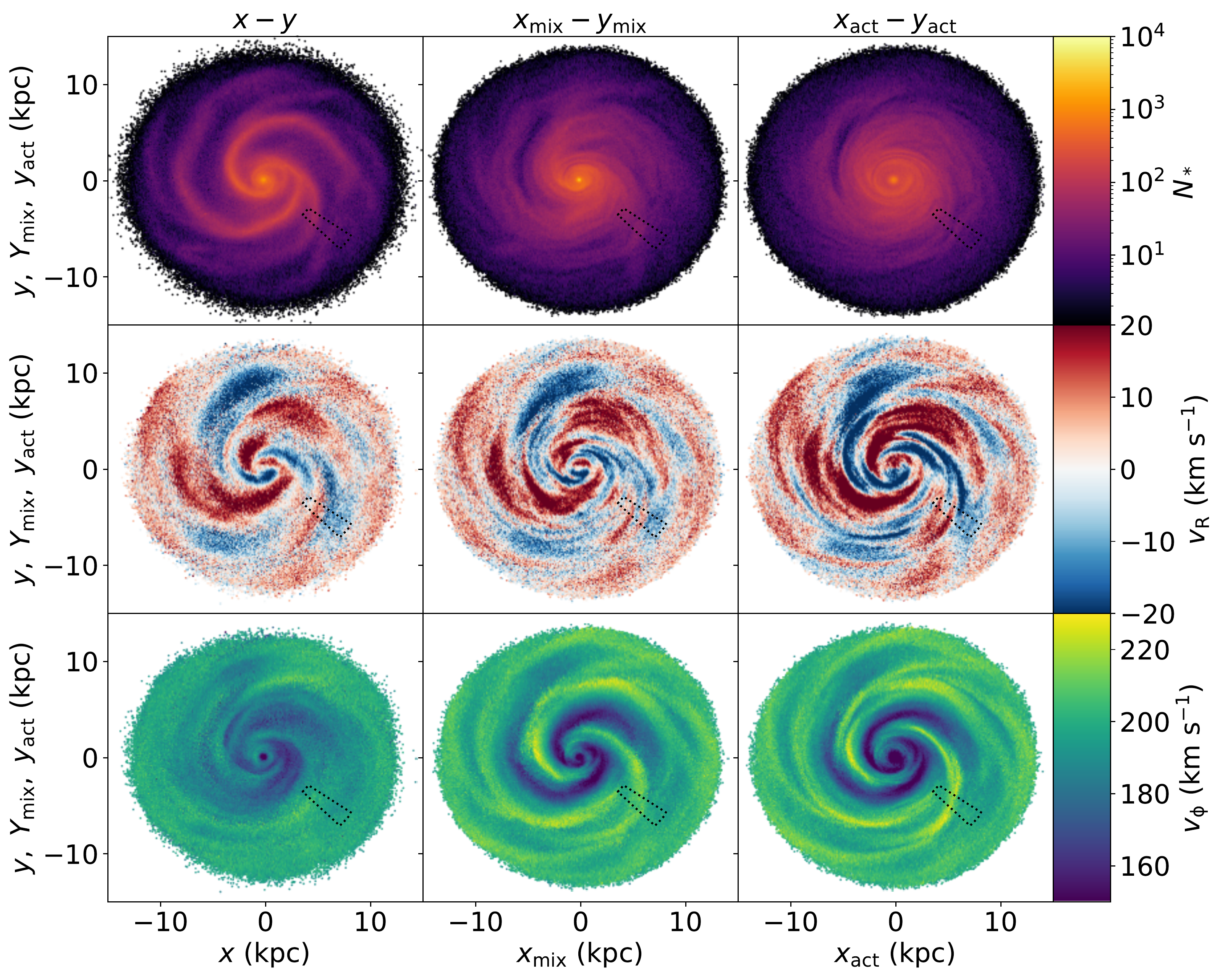}
    \caption{Same as Fig. \ref{K12} but for Model B. The dotted line shows the wedge selection of $315<\phi<325$ (deg) and $315<\theta_{\phi}<325$ (deg) used later in Fig. \ref{RadialWave}.}
    \label{AlexG}
\end{figure*}

\subsection{Barred galaxy (Model A)}
Firstly, we examine a pure $N$-body simulation run with \sc{GCD+ }\rm \citep[e.g.][]{KG03}, with $6\times10^7$ particles, which displays a strong bar, and some weak spiral structure, which henceforth we call Model A. The model is comprised of three disc components and a fixed NFW halo \citep[][]{NFW97} set up as described in \cite{GKC12}. The discs consist of a thin disc with $4.5\times10^7$ particles, a disc mass $M_{\mathrm{d},1}=4.5\times10^{10}$~M$_{\odot}$, a scale length $R_{\mathrm{d},1}={3.5}$ kpc, a scale height $z_{\mathrm{d},1}=0.25$ kpc, and $\sigma_{\mathrm{R}}^2/\sigma_z^2=4$, a thick disc with $10^7$ particles, $M_{\mathrm{d},2}=1\times10^{10}$~M$_{\odot}$, $R_{\mathrm{d},2}={2}$ kpc, $z_{\mathrm{d},2}=1$ kpc and $\sigma_{\mathrm{R}}^2/\sigma_z^2=1$, and another thick disc with $5\times10^6$ particles, $M_{\mathrm{d},3}=0.5\times10^{10}$~M$_{\odot}$, $R_{\mathrm{d},3}={3}$ kpc, $z_{\mathrm{d},3}=0.5$ kpc and $\sigma_{\mathrm{R}}^2/\sigma_z^2=1$. The NFW halo has a mass of $1.14\times10^{12}\ M_{\odot}$, and a concentration parameter of 14. Model A has $V_{\mathrm{circ}}(R=8.178)=200$ km s$^{-1}$.

Fig. \ref{K12} shows the face on view of Model A in logarithmic number density (upper row), mean radial velocity (middle row) and mean rotation velocity (lower row) for $x-y$ (left column), $x_{\mathrm{mix}}-y_{\mathrm{mix}}$ (middle column) and $x_{\mathrm{act}}-y_{\mathrm{act}}$ (right column). The left column shows the standard Cartesian representation of the model galaxy. The middle row shows that the bar has a strong quadrupole and the radial velocity signatures correlate with the spiral density enhancements in the top panel, but in a non-trivial way. The middle column shows Model A in guiding radius Cartesian coordinates, i.e. the half-transform, using the action but not the associated angle. Here, the model does appear to be significantly sharper, with numerous thin spiral features in the top panel, and more coherent velocity signatures in the middle and lower panels. However, these features in the density plot are not indicative of the current location of the spiral structure. The right column shows Model A in guiding centre Cartesian coordinates. Both the radial velocity and azimuthal velocity signatures become even stronger. However, there is more sharpening in the azimuthal velocity structure when moving from $x-y$ to $x_{\mathrm{mix}}-y_{\mathrm{mix}}$, and more sharpening in the radial velocity structure when moving from $x_{\mathrm{mix}}-y_{\mathrm{mix}}$ to $x_{\mathrm{act}}-y_{\mathrm{act}}$. This is to be expected, because the shift in radius is sorting stars by $L_z$, creating more coherent structure in $v_{\phi}$, and the shift in angle is sorting stars by $\theta_{\phi}$ creating more coherent structure in $v_{\mathrm{R}}$.  

Next, we select a small section of model A in a 1 kpc sphere around $(x,y)=(8.178,0)$ kpc, and transform this into guiding radius and guiding centre coordinates. The left part of Fig. \ref{sample} shows the sample in Cartesian coordinates (left column), guiding radius coordinates (centre column) and guiding centre coordinates (right column), for the logarithmic number density (upper row) and mean radial velocity (lower row). The left column shows the location of the selected sample with a black circle. The centre column shows a slightly tri-modal distribution in the number counts, and significant structure in the $v_{\mathrm{R}}$ map. The top panel of the right column shows a heavily structured distribution in the number density, with numerous features across many kpc which is not representative of the distribution in the left panel, and could resemble numerous thin spiral arms. The lower panel shows a clear split by $v_{\mathrm{R}}$ along the $y_{\mathrm{act}}=0$ line as shown for the data in Section \ref{data}. This effect is not seen when we have the full information for the galaxy (e.g. the middle row of Fig. \ref{K12}) but over a local volume stars with positive $v_{\mathrm{R}}$ will have their guiding centre azimuth further in the direction of rotation, and the opposite is true for stars with negative radial velocities.

Thus, to test the statement above, that the middle column is essentially $\phi-L_z$ and the right column is $\theta_{\phi}-L_z$, we plot these quantities in the right part of Fig. \ref{sample}, which shows the same sample as a function of angular momentum against galactic azimuth $\phi$ (left column) and guiding centre azimuth (right column), for the logarithmic number density (upper row) and colored by radial velocity (lower row). A comparison of the left and right parts of Fig. \ref{sample} show the distributions to be extremely similar, with the only difference being the curvature of the frame away from the $y_{\mathrm{mix}}=0$ or $y_{\mathrm{act}}=0$ line. As discussed above, the middle column of the left part of Fig. \ref{sample}, is mixing coordinate systems. While this blend of galactic azimuth and angular momentum can be explored, and can highlight interesting dynamical phenomena in the disc \citep[e.g.][]{FS19}, we argue that the full transformation provides more information on particles, or stars, with shared orbital histories as illustrated in the right column of the left part of Fig. \ref{sample}. However, for such a small volume, the selection function is extremely important, and care must be taken in the interpretation. Note also that this illustration is a very simplistic selection function when compared with data such as from $Gaia$.


\subsection{Spiral galaxy (Model B)}
Secondly, we examine a $N$-body/smoothed particle hydrodynamics simulation of a spiral galaxy that lacks a central bar, allowing for a focused analysis of arm features, which we call Model B. This is a reproduction of the Bc Milky Way model from \citet{Pettitt2015}, wherein several different disc configurations were explored to find the best reproduction of the spiral features seen in the \citet{DHT01} CO longitude-velocity Milky Way data. This model produced a moderate amount of spiral structure while a large bulge component prevented the formation of a bar. This specific version has been reproduced using the \textsc{Gasoline} code \citep{Wadsley2017}, and contains a live stellar disc, bulge and gas disc embedded in an inert dark matter halo. Unlike \citet{Pettitt2015}, this version includes self-gravity in the gas and star formation, cooling and feedback subgrid physics (following \citealt{Keller2014}). The simulation initially contains $2\times 10^6$ gas and $2\times 10^6$ stellar disc and $1\times 10^5$ bulge particles, but over time gas particles are converted into stars, providing 2,522,592 star particles at the time of the snapshot. See \citet{Pettitt2015} for details on the initialisation procedure. Due to these additional physics the galactic morphology is not expected to be identical to the data presented in \citet{Pettitt2015}. An arbitrary snapshot was chosen that displayed strong spiral features, and is not meant to be an exact reproduction of the Milky Way's spiral structure which is expected to be weaker.

Fig. \ref{AlexG} shows the same as Fig. \ref{K12}, except for Model B, a mostly grand design spiral galaxy model where the spiral arms arise purely from the disc instability. There is no bar in this model, and thus we do not need to decompose the effect of the bar resonances and the transient spiral arms. The middle and right hand columns show that the transformation to guiding radius, and guiding centre coordinates does not make the spiral arms more clear in the density (upper row), but rather the reverse, breaking the single strong spiral arms into multiple thin features. The middle and lower rows of Fig. \ref{AlexG} show again that the kinematic features become more coherent with the transformation to guiding coordinates.

\begin{figure}
    \centering
    \includegraphics[width=\hsize]{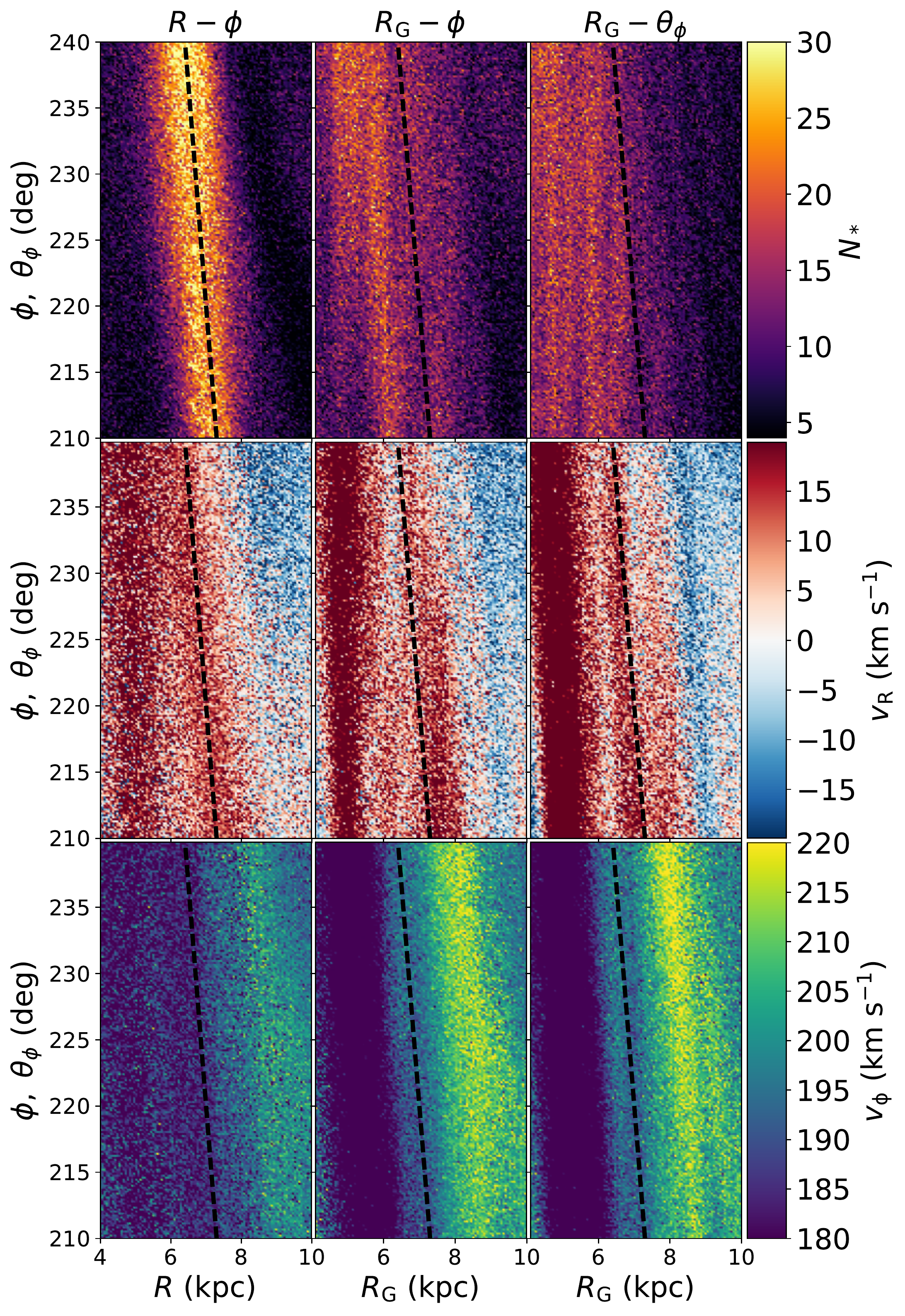}
    \caption{Model B projected into the $R-\phi$ plane for a region with $210<\phi<240$ deg, and $|b|<5$ deg, for standard Cartesian coordinates (left column), guiding radius Cartesian coordinates (centre column) and $210<\theta_{\phi}<240$ deg for guiding centre Cartesian coordinates (right column), for surface density (upper row), mean radial velocity (middle row) and rotation velocity (lower row). The dashed lines mark the centre of the density enhancement in physical space. Note that for this visualisation only, we have stacked five snapshots spaced by 1 Myr to increase the effective resolution.}
    \label{armrv}
\end{figure}

To further illustrate this, Fig. \ref{armrv} shows the $R-\phi$ plane (left column), the $R_{\mathrm{G}}-\phi$ plane (centre column) and the $R_{\mathrm{G}}-\theta_{\phi}$ plane (right column), for surface density (upper row), mean radial velocity (middle row) and mean rotation velocity (lower row), over a small region of the model chosen to isolate a spiral arm slightly inside the Solar radius. For visualisation purposes only, we have stacked five snapshots spaced by 1 Myr and rotated to match the position angle in order to increase the effective resolution. Note that the dynamics change little on such a timescale and the observed patterns are consistent across individual snapshots.

The upper left panel of Figure \ref{armrv} shows a segment of a single strong spiral arm in the number density, with the centre of this density enhancement overlaid in all panels (dashed line). The middle left and lower left panels show kinematics that are clearly correlated with the spiral arm. The top middle and top right panels show that as seen in Fig. \ref{AlexG} the spiral arm is smeared out into multiple sub-components in both the guiding spaces. This makes sense, because while the stars which are currently located in the density enhancement of the spiral arm may be co-spatial in physical space, they are not all on the same orbit, and thus, once we perform the coordinate transform to actions and angles, which are essentially orbit labels, it is unsurprising that stars on similar orbits become grouped, rather than those currently in the spiral density enhancement. As already discussed above, for the guiding radius space (middle column), the rotation velocity signature becomes sharper than in $R-\phi$ because stars are sorted by their angular momentum, whereas the radial velocity signature shows little change. Similarly, when completing the transformation to the Guiding centre Cartesian coordinates (right column) the radial velocity signature becomes much sharper because stars are now being sorted by orbital angle, whereas the rotation velocity shows little change compared to the guiding radius space.

There is no reason that we cannot examine the galactic kinematics in these spaces, but the ridges in $x_{\mathrm{mix}}-y_{\mathrm{mix}}$ or $x_{\mathrm{act}}-y_{\mathrm{act}}$ do not correspond to the present overdensities of the Milky Way spiral structure, and instead group stars on shared orbits as discussed above. The top left panel of Figures \ref{AlexG} and \ref{armrv} show a representation of the structure, and potential, but their right hand columns show the response to the potential, not the potential itself. The middle column shows a mix of two canonical coordinate systems, and care should be taken in the interpretation, for it is not entirely the potential, nor the response to the potential.

Again, if we wish to examine the current distribution of mass, we should be looking in physical space, e.g. the top left panel of Figures \ref{AlexG} and \ref{armrv}, but if we wish to examine the kinematic response to the potential we should be looking in action-angle space, e.g. the middle right and lower right panels of Figures \ref{AlexG} and \ref{armrv}. Even if we had $Gaia$ data for the whole of the Milky Way, we would be better off visualising the density enhancement in the original frame. We can, and should, use the kinematic response to inform our knowledge of the potential, but it is not a direct map.

\begin{figure}
    \centering
    \includegraphics[width=\hsize]{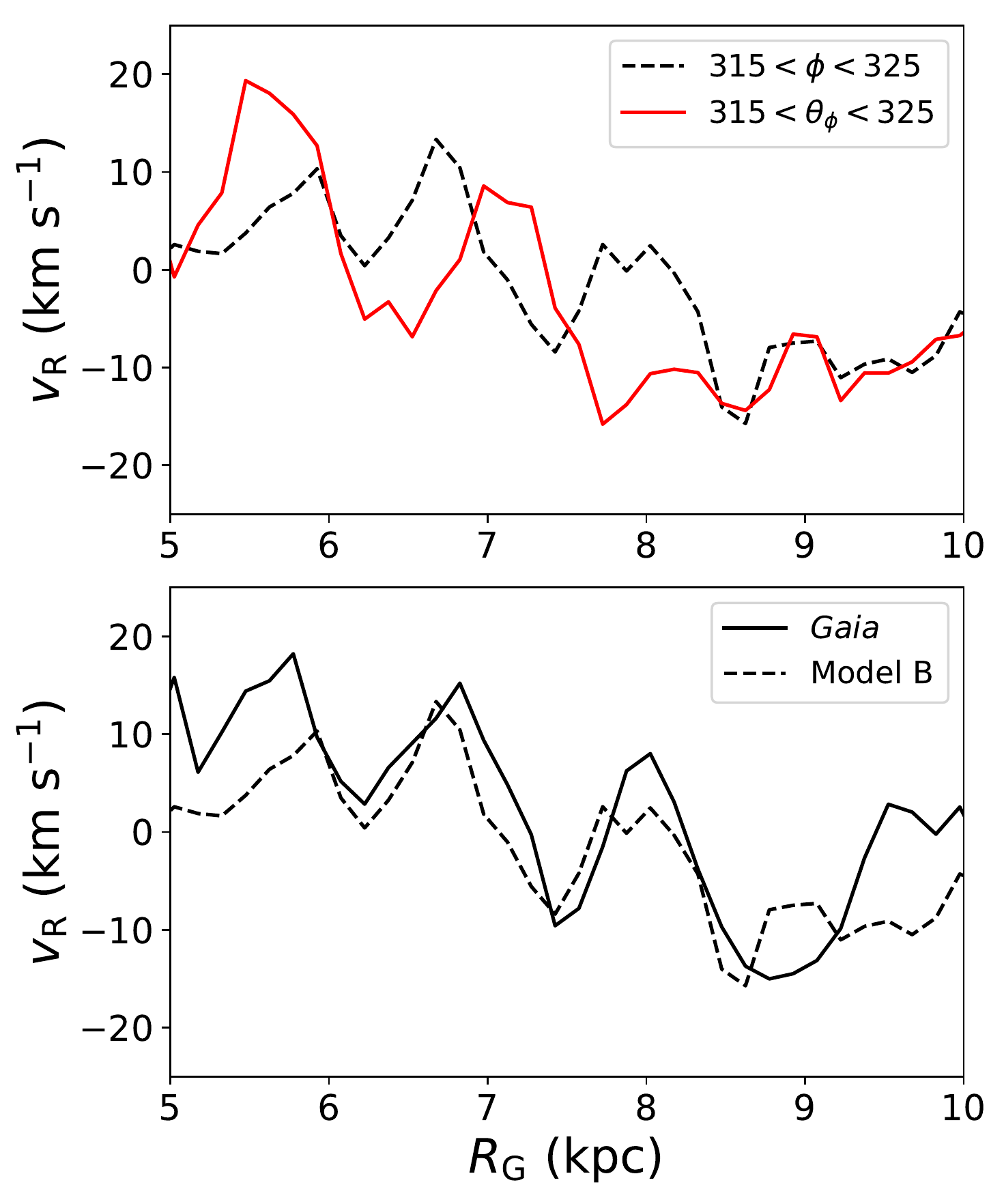}
    \caption{\textbf{Upper:} Mean radial velocity as a function of guiding radius along a $315<\phi<325$ (deg) wedge (black dashed), and along a $315<\theta_{\phi}<325$ (deg) wedge (red solid), with $\mid b \mid<5$ (deg) from Model B. \textbf{Lower:} $R_{\mathrm{G}}-v_{\mathrm{R}}$ in the $Gaia$ data along a thin wedge with $-1<\phi<1$, $\mid b \mid<1$ deg (solid) overlaid with Model B shown along the same wedge as the upper panel, with $315<\phi<325$, $\mid b \mid<5$ deg (dashed).}
    \label{RadialWave}
\end{figure}

In addition, while in the example shown in Figure \ref{armrv} the pitch angle of the physical spiral in the top left panel is larger than the pitch angle of the features in the top middle and top right panels. The true pitch angle does match the pitch angle of the velocity signatures in the lower rows, but this was not always the case in other spiral models we examined.

The angle of such features is easily affected by spurs, resonances, or the influence of external perturbers making it an unreliable indication of spiral pitch angle. For example, the angle of kinematic features which arise from resonances is determined by the order of the resonance \citep[e.g. as discussed in relation to the $Gaia$ data in][]{Monari+19,FS19}, and the angle of the kinematic features arising from interaction with a satellite is dependent on the time-scale of the phase mixing.

A detailed examination of the differences between the angles of features in this space is beyond the scope of this work. However, there is currently significant disagreement on the origin of the kinematic substructure in the Solar neighbourhood, and as such we do not consider it wise to measure spiral pitch angle from features that may arise from one of several different causes.

\section{The radial velocity wave}\label{wave}
Numerous recent works have examined signatures in the Galactocentric radial velocities of stars in $Gaia$ DR2, both in the $R-v_{\phi}$ plane \citep[e.g.][]{Fragkoudi+19,Hunt+19}, and the $L_z-\phi$ (or $R_{\mathrm{G}}-\phi$) plane \citep{FS19}. A wave-like pattern has also been observed in the Galactocentric radial velocities, both locally in the $Gaia$ data by \cite{FS19} (which is also strongest as a function of guiding radius, see their Fig. 7), and globally in the APOGEE data by \cite{Eilers+20}, who find a wave pattern in the radial velocity signatures over a much larger range of the disc. The local part of this large scale radial velocity wave has been previously observed as a radial velocity gradient \citep[e.g.][]{Siebert+2011,Williams+2013} in data from the Radial Velocity Experiment \citep[RAVE;][]{Sea06}, although they lacked the spatial coverage to resolve the whole waveform.

However, as shown in the middle row of Fig. \ref{armrv} kinematic signatures should becomes stronger when completing the transformation to action angle coordinates. Thus, the radial velocity wave should also be stronger in $R_{\mathrm{G}}-\theta_{\phi}$ than $R_{\mathrm{G}}-\phi$. The top panel of Fig. \ref{RadialWave} shows the mean radial velocity as a function of guiding radius along a $315<\phi<325$ (deg) wedge (black dashed), and along a $315<\theta_{\phi}<325$ (deg) wedge (red solid), with $\mid b \mid<5$ (deg) from Model B. 

The amplitude of the wave like signature is larger in $R_{\mathrm{G}}-\theta_{\phi}$ than in $R_{\mathrm{G}}-\phi$ as expected. In this model, the wave signature arises from transient spiral structure. However, it is not surprising that transient winding arms create such a signature in $R_{\mathrm{G}}-v_{\mathrm{R}}$ because it has already been shown in $R-v_{\phi}$ \citep[e.g.][]{Hunt+19,Khanna+19}, and this is merely a different projection of the kinematics of stars around transient winding spirals examined in multiple works \citep[e.g.][]{GKC14,KHGPC14}.

The lower panel of Fig. \ref{RadialWave} shows $R_{\mathrm{G}}-v_{\mathrm{R}}$ in the $Gaia$ data \citep[as previously shown in][]{FS19} along a thin wedge with $-1<\phi<1$, $\mid b \mid<1$ deg (solid) overlaid with Model B shown along the same wedge as the upper panel, with $315<\phi<325$, $\mid b \mid<5$ deg (dashed). Note that this is not intended to be a best fit to the wave pattern as Model B is not fit to the Milky Way. However, we note that multiple lines of sight in Model B produce a qualitative match to the $Gaia$ data providing the observer is slightly outside a spiral arm. The model does not recover the outermost peak around $R_{\mathrm{G}}=10$ kpc, but this is likely explained by the lack of a Perseus like arm in this snapshot.

While \cite{FS19} show that the wave is stronger in $R_{\mathrm{G}}-\phi$ than $R-{\phi}$, they do not examine the $R_{\mathrm{G}}-\theta_{\phi}$ plane, which would currently be challenging given the observational selection effects as discussed in Sections \ref{grad} and \ref{simulation}. As discussed above, over a local sample, stars with guiding centre azimuths close to zero, also have radial velocities close to zero. However, with future $Gaia$ data releases, and other upcoming surveys with larger spatial coverage, we may be able to learn more by completing the transformation.

We examined multiple other $N$-body spiral galaxy models with varying morphology, and determined that while all of them produce a wave in the radial velocities, the wavelength, amplitude and regularity vary significantly between models and the line-of-sight chosen. Further study would be needed to determine how the parameters of the spiral arms affect the resulting wave, but that is beyond the scope of this work \citep[although see][for a fit of a steady state model to the global wave pattern]{Eilers+20}. Further study would also be needed to determine whether such a signal is different for spiral structure generated with different underlying theories of spiral arm formation. For example, \cite{FS19} suggest a link between the wave in the radial and vertical kinematics which would not be expected if it is caused by a transient spiral or a density wave arising in an isolated disc. However, if such a spiral has been generated via the interaction with a satellite perturber, it may be natural to expect such signals to correlate \citep{LMJG19}.  

It should be possible to fit such a model of the response to the transient spiral structure to both the local detailed wave pattern from \cite{FS19} and the global pattern as done for a steady state model in \cite{Eilers+20}. However, the galactic bar and interactions with satellites have undoubtedly also shaped such a feature, making it hard to draw a firm conclusion. Thus we defer a more detailed exploration to future work, and for now conclude merely that transient spiral structure can naturally produce the radial velocity wave feature observed in the Solar neighbourhood, and beyond.

\section{Summary}\label{summary}
In this work we have attempted to illustrate the power of actions and angles in resolving the kinematic structure of the Solar neighbourhood and further across the Galactic disc, while also providing an illustration of the impact of the selection function, and a warning on the interpretation of data when mixing canonical coordinate systems. Our conclusions are summarised as follows:

1) We advocate the use of two distinct projections to physically explore the disc data. Firstly, in the projected physical space the overdensities trace the potential, and the velocities trace scattering. Secondly, in the projected action-angle space the overdensities trace stars on common orbits. Orbits with common frequencies have common orbit averaged reactions, and orbits with common frequencies and common angles have shared history and preserve the dynamical memory of the past. 

2) We caution against mixing the coordinate systems because this obscures the interpretation of the data. As an example, we have shown that the numerous ridges in guiding radius and guiding centre space, observed in the $Gaia$ DR2 data \citep{Khoperskov+19b} are not physical density enhancements corresponding to spiral arms, but rather groups of comoving stars with a shared orbital history which have long been known in the Solar neighbourhood. The guiding centre space, or action-angle space provides a good frame for us to examine the kinematic response to the Galactic potential, but it is not a direct map of the current Galactic spiral structure. For the moment, we can certainly learn from the mixed coordinate system \citep[e.g.][]{FS19} while selection effects prevent us from fully utilizing the angles, but we argue that once we have a more global view of the galaxy we should be using the corresponding conjugate angle when analysing actions.

3) We find that the global disc responses are clarified using the action-angle space, and we continue to show that for a model of transient winding spiral structure this space splits stars with different orbital history into different streams, and that a coherent spiral arm in physical space is not a single structure in $x_{\mathrm{mix}}-y_{\mathrm{mix}}$ or $x_{\mathrm{act}}-y_{\mathrm{act}}$. We then link these multiple features with the wave in the radial velocities shown locally by \cite{FS19}, and globally by \cite{Eilers+20}. While such a wave should be stronger as a function of $R_{\mathrm{G}}-\theta_{\phi}$ than $R_{\mathrm{G}}-\phi$ (or $R-\phi$), selection effects currently make it difficult to examine this space in the data. Whether we can fit exactly the wave pattern to a combination of the Galactic bar and spiral structure is beyond the scope of this work, but we show here that transient spiral arms naturally produce such a wave in the radial velocities.

Overall, we conclude that this will become increasingly important as we move towards a global view of the Milky Way, enabling us to fully take advantage of the angles. For the moment, working in the mixed coordinate systems can tell us more than positions and velocities alone, but care must be taken in the interpretation. We look forward to the future $Gaia$ data releases, in combination with the next generation of ground based spectroscopic surveys such as SDSS-V Milky Way Mapper.

\section*{Acknowledgements} The authors thank the anonymous referee for a constructive report. JASH \& ECC are supported by a Flatiron Research Fellowship at the Flatiron institute, which is supported by the Simons Foundation. KVJ was supported by NSF grant AST-1715582. DK acknowledge the support of the UK's Science and Technology Facilities Council (STFC Grant ST/K000977/1 and ST/N000811/1). This research made use of \texttt{astropy}, a community-developed core Python package for Astronomy \citep{Astropy}, and the galactic dynamics Python package \texttt{galpy} \citep{B15}. This work has also made use of data from the European Space Agency (ESA) mission $Gaia$ (https://www.cosmos.esa.int/gaia), processed by the $Gaia$ Data Processing and Analysis Consortium (DPAC, https://www.cosmos.esa.int/web/gaia/dpac/consortium). Funding for the DPAC has been provided by national institutions, in particular the institutions participating in the $Gaia$ Multilateral Agreement. Numerical computations were (in part) carried out on Cray XC50 at Center for Computational Astrophysics, National Astronomical Observatory of Japan.

\bibliographystyle{mn2e}
\bibliography{ref2}

\label{lastpage}
\end{document}